\documentclass[a4paper,11pt]{article}

\usepackage{caption}
\usepackage{subcaption}
\usepackage{pos}
\usepackage[super]{nth}
\usepackage{graphicx}
\usepackage{wrapfig,lipsum,booktabs}

\title{Tidal Disruption:  An Unforgettable Encounter with a Black Hole}

\ShortTitle{Unforgettable TDE Encounter}

\author{The IceCube Collaboration \\{\normalsize \normalfont(a complete list of authors can be found at the end of the proceedings)}\\}

\emailAdd{mdhosale@yorku.ca}
\emailAdd{madsen@icecube.wisc.edu}
\emailAdd{vbasu@wisc.edu}

\abstract{

The annual GLEAM outdoor art exhibition features curated, large-scale light installations in the Olbrich Botanical Gardens in Madison, Wisconsin. With submissions from local, regional, and international artists, the annual two-month long event draws tens of thousands of visitors each fall. “Tidal Disruption,” an art-science light sculpture representative of the form and behavior of a tidal disruption event, debuted at GLEAM 2022. The installation was a collaboration between artists Hosale and Abdu’Allah, and astrophysicist Madsen. The immersive light and sound exhibit stylistically depicted the death of a star as it spirals into a black hole in a roughly 4 minute sequence. Designed primarily by Hosale, the components were custom fabricated using clear PVC pipe at the Physical Sciences Laboratory at UW-Madison. In this poster, we will describe the process of developing, fabricating, and installing “Tidal Disruption” and the response from the viewers.

\vspace{4mm}
{\bfseries Corresponding authors:}
Mark-David Hosale$^{1}$, Jim Madsen$^{2}$, Vedant Basu $^{2*}$ \\
{$^{1}$ \itshape Computational Arts, School of Art, Media, Performance and Design
York University, Toronto}\\
{$^{2}$ \itshape WIPAC, UW-Madison, USA}\\
[4mm]
$^*$ Presenter

\ConferenceLogo{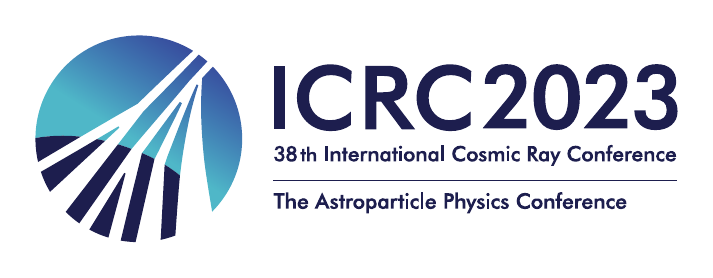}

\FullConference{The 38th International Cosmic Ray Conference (ICRC2023)\\ 26 July -- 3 August, 2023\\ Nagoya, Japan}
}

\begin{document}

\maketitle

\section{Introduction}\label{sec1}
The IceCube Collaboration continues to seek new ways to engage with diverse audiences.  One path to success in this regard is to partner with artists whose works convey science from a different perspective, providing access to communities that are less likely to attend a more traditional IceCube outreach event.  A few examples of prior work with artists have been described in previous ICRC meetings, \cite{Madsen:20194t, Madsen:2021D9}. 
Here we describe an ambitious effort to convey the catastrophic process of a star being destroyed by a black hole, referred to as Tidal Disruption Event (TDE).

For many years, Olbrich Botanical Gardens in Madison, Wisconsin has hosted an annual exhibition featuring artistic light installations, \href{https://www.olbrichgleam.org/}{GLEAM}.  Preliminary proposals are due about a year in advance of the show, of which a few are selected to submit full proposals.  From these, approximately ten projects are selected with a budget up to \$10,000.00 including in-kind contributions.  Installation takes place in early August, and the shows are open to the public Wednesday-Saturday evenings in September and October. An admission fee  helps fund the projects and support operation of Olbrich Gardens. The total attendance of approximately 30,000 people is much higher than typically drawn for science installations on the UW-Madison campus.  

\section{Concept Development}\label{concept}
\begin{figure}[b]
    \centering
    \subfloat[\centering]{{\includegraphics[width=6.85cm]{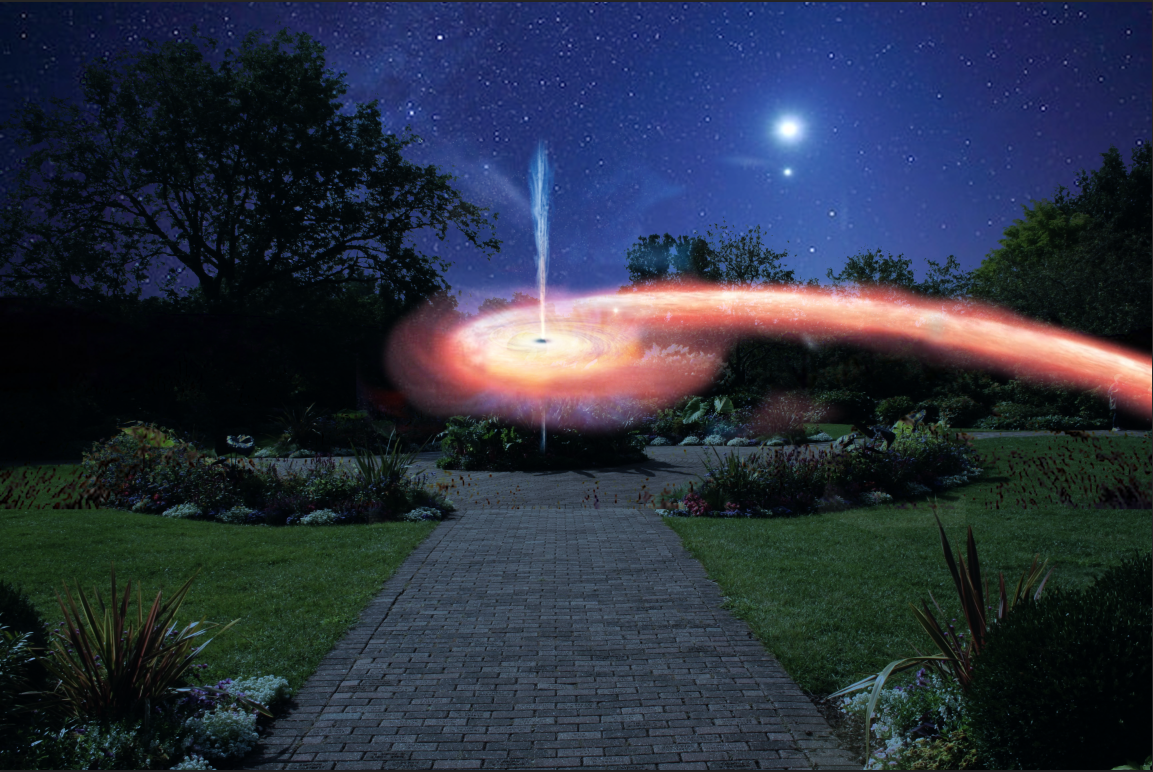}}\label{fig:vision1}}
    \qquad
    \subfloat[\centering]{{\includegraphics[width=7.35cm]{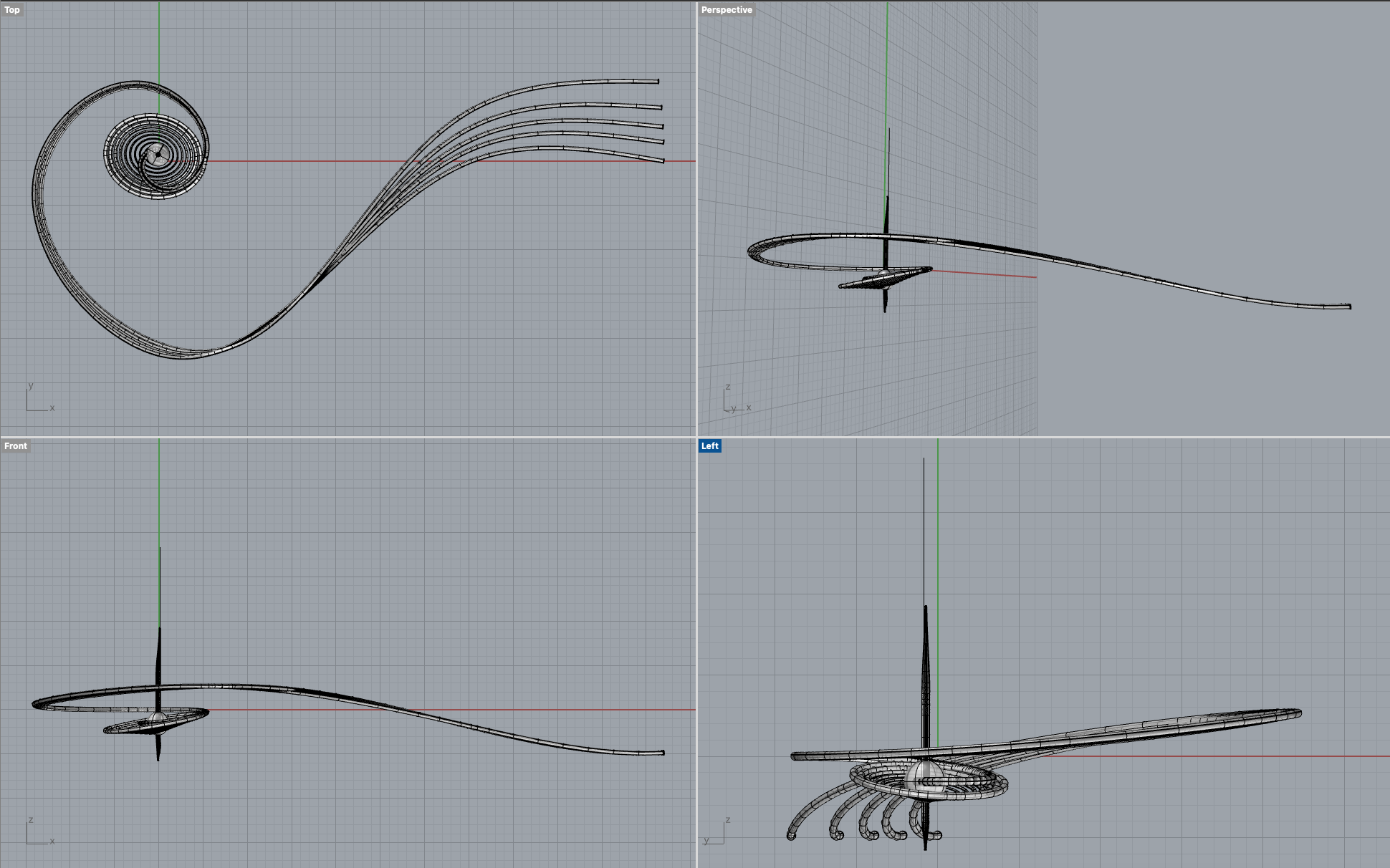}}\label{fig:wireframe}}
    \caption{\textbf{(a)} A first artistic vision of the TDE light sculpture. \textbf{(b)} Preliminary wire frame rendering.}
    \end{figure}

A preliminary discussion led to a commitment to put together a proposal, and an initial design idea was developed by Hosale, see Figure~ (\ref{fig:vision1}). The preliminary proposal was crafted in parallel.

\textbf{Installation Synopsis:}
Massive black holes lurk at the center of galaxies, gobbling wayward stars that wander too close. What once was a beautiful ball of fusing matter is torn apart by the phenomenal gravitational strength of the unimaginably dense black hole.
Viewers would set off a light and sound “explosion,” representative of the explosive jet of high-energy particles produced when the tidal forces become so strong that not even a star can survive.

An architectonic light sculpture representative of the form and behavior of a tidal disruption event as described above was designed, executed, and installed. 
Viewers experienced the cosmic fireworks produced by the encounter as they walked around an interactive model of the black hole. 
The overall experience aimed to express the sublime nature of black holes, exposing the mysterious beauty and power of cataclysmic cosmic events in a thought-provoking installation where art and science collide.

The next step was to move from conceptual design to a wire frame model (Fig.~\ref{fig:wireframe}). 

\begin{figure}[h!]
  \begin{center}
  \includegraphics[width=5.5in]{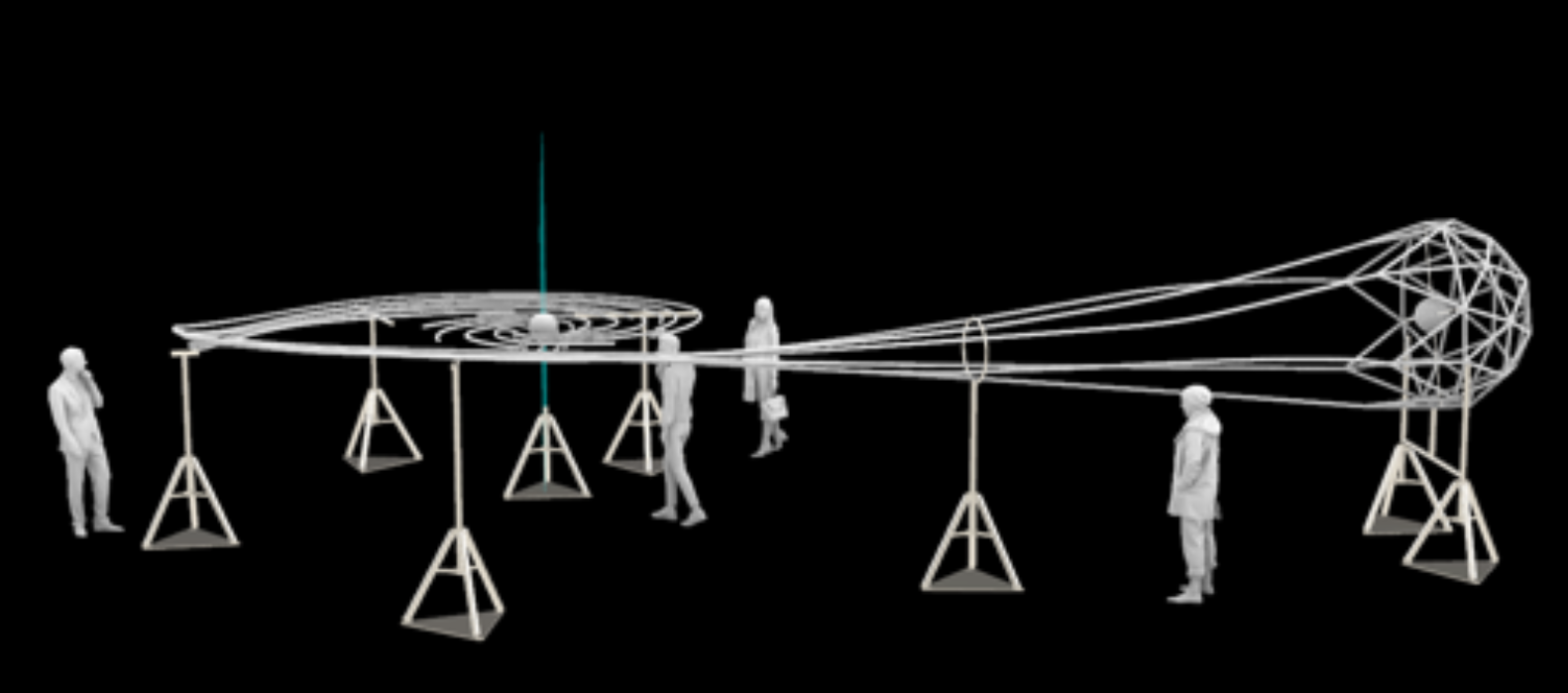}
  \caption{The final rendering of the TDE light sculpture.}
  \label{fig:wireframe2}
   \end{center} 
\end{figure}

This model progressed to a buildable design (Fig. ~\ref{fig:wireframe2}),  that needed to satisfy design goals and constraints described below. 

\textbf{Dimensions:} Approximately 60' in length and 20' at its widest point.

\textbf{Materials:}
The main body of the sculpture was constructed of clear PVC tubing
illuminated with internal LEDs. The black hole was represented by a black sphere with acrylic rods from the top and bottom that could be “energized” by pulsing a bright light inside the sphere to represent particle jets. 
The PVC tubing was supported by a structure painted black to minimize visibility at night.
Earth anchors screwed into the ground secured the work.

\begin{figure}[h!]
    \centering
     \subfloat[\centering]{{\includegraphics[width=5.5cm]{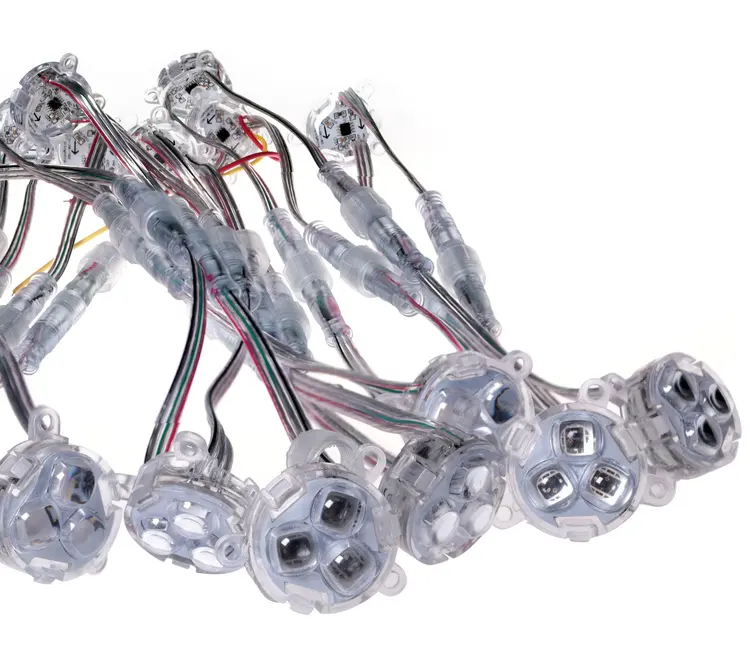}}\label{fig:leds}}
       \qquad
     \subfloat[\centering]{{\includegraphics[width=7.5cm]{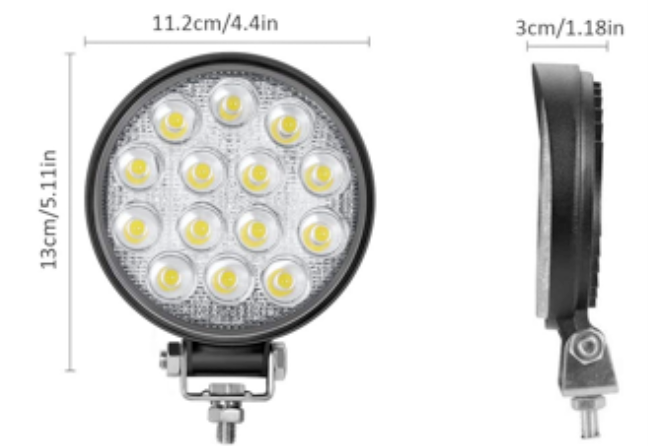}}\label{fig:headlight}}
    \caption{\textbf{(a)} The 3-element LED disks used to illuminate the clear PVC pipes and \textbf{(b)} the off-road headlights used to light the star and black hole jet.}
    \end{figure}

\begin{figure}[b]
    \centering
     \subfloat[\centering]{{\includegraphics[width=8.35cm]{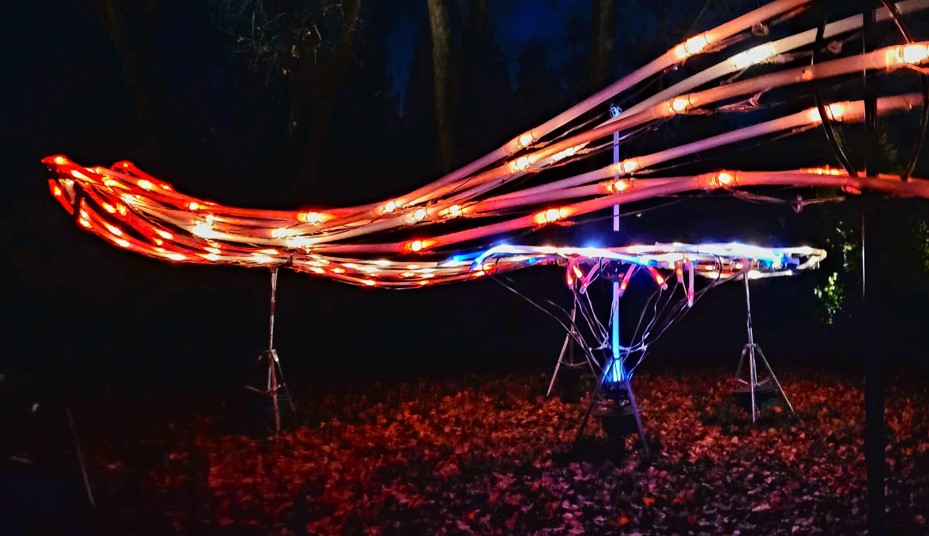}}\label{fig:bhole}}
       \qquad
       \subfloat[\centering]{{\includegraphics[width=5.65cm]{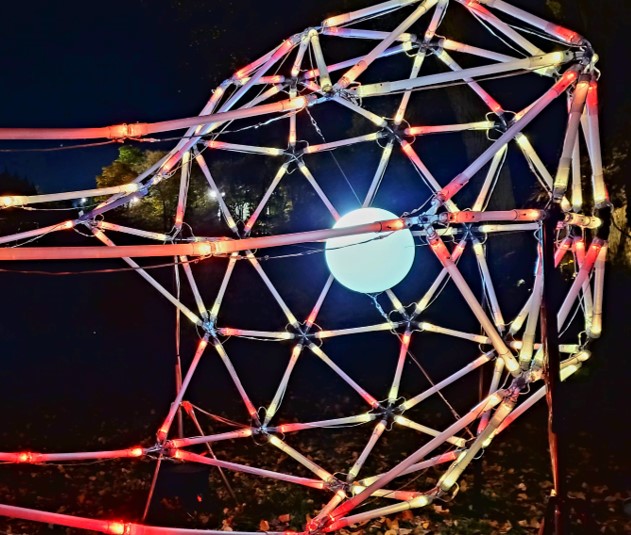}}\label{fig:Closestar}}
    \caption{Close ups of the black hole, \textbf{(a)}, and star, \textbf{(b)}. }
    \end{figure}

\textbf{Lighting Fixture Description 
:}
The entire installation was lit using three strategies:
\begin{enumerate}
    \item {Side emitting clear PVC pipes illuminated internally form the main bodu of the work.}
\item{Commercially available \href{https://www.shiji-led.com/led-pixel-light/sj-30-ucs1903.htm}{computer-controllable strings of LED discs}, Fig.~\ref{fig:leds}, show the
progression of matter from the star being pulled into the black hole.}
\item{Higher power computer-controllable LEDs, Fig.~\ref{fig:headlight}, illuminated the globe at the center of the star, Fig.~\ref{fig:Closestar} and the vertical acrylic sculptural rods that
represented the jets of particles emanating from the axis of the black hole, Fig.~\ref{fig:bhole}.}
\end{enumerate}
In a steady state, the installation cycled from white light to reddish and orange light funneling from the outer end
of the sculpture, spiraling and accelerating into the black hole, Figs.~\ref{fig:wlight}-\ref{fig:fullcolor}.

\begin{figure}[]
  \begin{center}
  \includegraphics[width=5.8in]{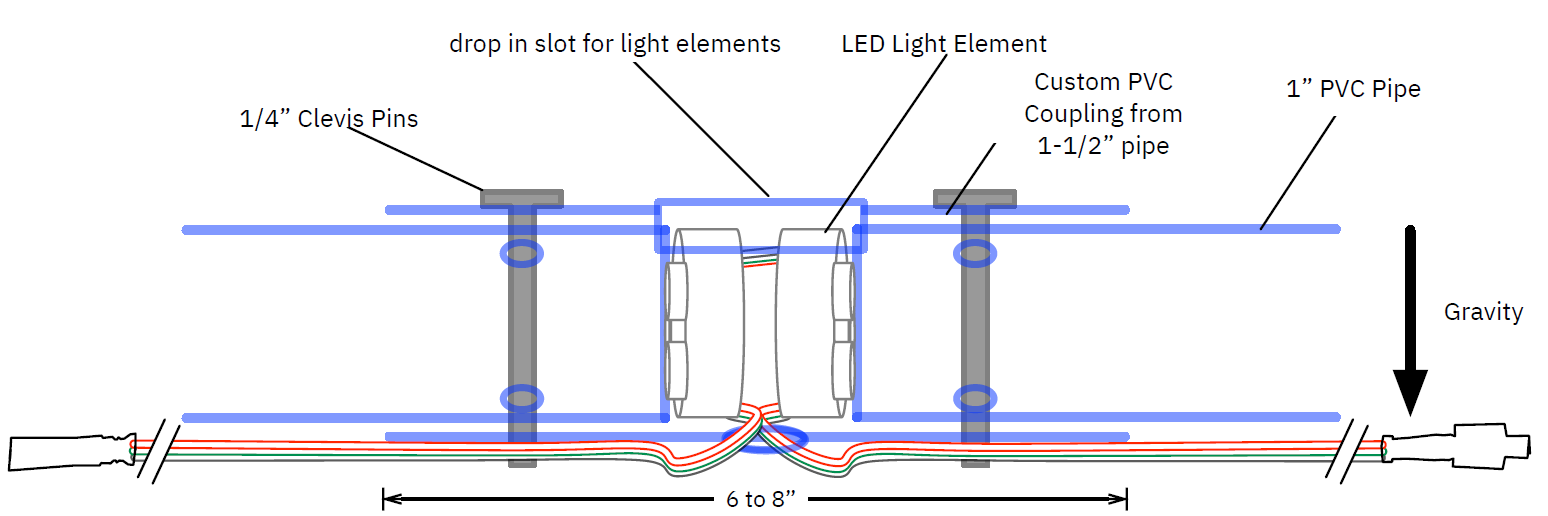}
  \caption{The the LED placement and mechanical connection between PVC pipes.}
  \label{fig:PVC}
   \end{center} 
\end{figure}

\textbf{Safety structures:}
The goal was to produce a light structure appearing suspended in space. We achieved this by have main supports at the star and black hole, with the light tubes representing matter streaming into the black hole held up by tripods as needed.
The earth anchors provided added stability. The
design also needed to:
\begin{itemize}
    \item {discourage but survive leaning on, shaking, and limited climbing;}
\item{withstand harsh conditions like high winds and freeze-thaw cycles; and}
\item{minimize site impact by making everything self-supporting.}
\end{itemize}

\textbf{Power connection points and total power:}
A power drop of 1000 watts was required. Power and all light controls were located under the black hole ball in the cantilever portion of the structure and intermediate support structures. 

\section{Design, Application, and Physical Layers}\label{application_and_physical}

The form of the installation was developed by Hosale using McNeel's Rhinoceros \cite{Rhino:McNeel}, a 3D modelling software that integrates well with digital fabrication workflows and is commonly used in architectural scale design. As alluded to above, the flexibility of the software allowed the creation process to move from line drawing sketches to full 3D rendering of every component and fastener in a single environment. 

Puck-like LEDs that were specifically designed for outdoor applications were used in the installation, Figure~\ref{fig:leds}. The wire lengths and wiring of the LED's were custom fabricated by a company in China \cite{ShijiLighting} with exact lengths calculated based on the 3D rendering provided. Lights were then strung along the PVC pipe with LEDs dropped in at regular intervals of approximately two feet along the length of the installation. At each connection point two LEDs were placed back to back to illuminate the pipe from both ends. Power was injected into each line of LEDs from power supplies in the tripod supports. 

The color of the LEDs were controlled using the same protocol that is used to control the widely used Neopixel built around the WS2812 chip and its variants \cite{Neopixels}. This allowed the installation to be controlled using readily available libraries found in the Arduino IDE microcontroller programming environment \cite{Arduino}. The physical layer of the LED control consisted of a Teensy 4.0 \cite{Teensy} with a OctoWS2811 Adaptor \cite{OctoWS2811} that allowed the installation to be addressed in a similar manner as a video wall. 

The LED control application was developed in C++ using OpenFrameworks \cite{OpenFrameworks}, a multi-platform C++ framework for computational based artworks. This application functioned as a simulator of the project as well as a control application for the final installation. The simulation mapped colors on a 3D model of the installation, facilitating the composing of light and sound behaviors before the installation was even built. The control application was deployed on a Raspberry Pi 3 Model B+ \cite{RaspberryPi} a single board low power computer that was connected to the Teensy with OctoWS2811 Adaptor and housed in a tripod at supporting the star portion of the installation. 

The behavior of the work was controlled by a manager application running on another Raspberry Pi in the other tripod support of the star. The manager application was developed in a computer music software called SuperCollider \cite{SuperCollider}, and took advantage of its musical timing features to control the sound of lights of the installation via network messages sent to RaspberryPis housed in each tripod of the installation. 

Each of the eight tripods , including the one used for light control, had a Raspberry Pi and a sound system. Each Raspberry Pi ran a SuperCollider application that played real-time generated soundscapes that were controlled via messages from the manager application, as described above. This allowed sound to spatially propagate through the work as a sonic expression of the process of the black hole eating the star.

\section{Fabrication}\label{fabrication}
The bulk of the fabrication work took about three weeks and involved cutting a total 284 pieces of clear 1.5" diameter PVC pipe in lengths that ranged from 9" to 34.5".  Most of these pieces had to subsequently be heated and bent to a predetermined arc. Finally, their surfaces were roughened with coarse sand paper to ensure diffusive light scattering along the length of the pipe. Larger diameter pipes were cut into 6" long sections that were used as connectors. Cutting, bending, drilling and sanding took about 120 people-hours.  

At each juncture of the star stream, LED "pucks" (Figure~\ref{fig:leds}) were placed back-to-back held in place by the clear PVC fittings and then fastened with clevis pins (which allow for quick installation and deinstallation), Figure~\ref{fig:PVC}. For the star, 3, 4, or 5 armed connectors were fabricated that provided both mechanical support and held a single LED "puck" in place. 

Tripods were configured from readily available antenna supports. These were bolted to plywood bases painted black. A shelf was added to hold the electronics which were contained in molded plastic boxes.  Uprights and cross pieces were cut from 2" PVC pipe painted black.  Electrical conduit tape was used to reinforce the PVC as necessary. 

\begin{figure}[]
    \centering
     \subfloat[\centering]{{\includegraphics[width=7cm]{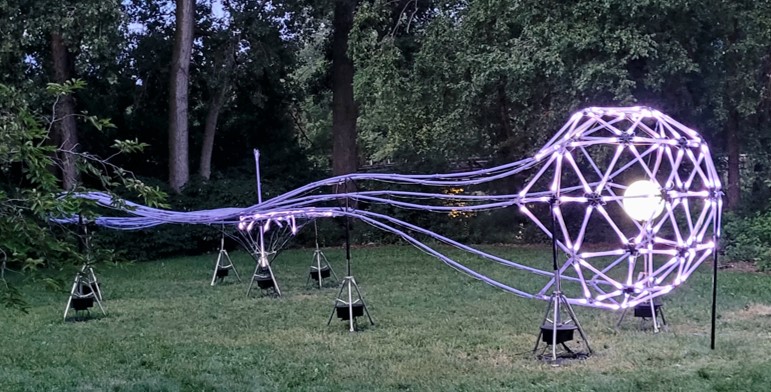}}\label{fig:wlight}}
       \qquad
     \subfloat[\centering]{{\includegraphics[width=7cm]{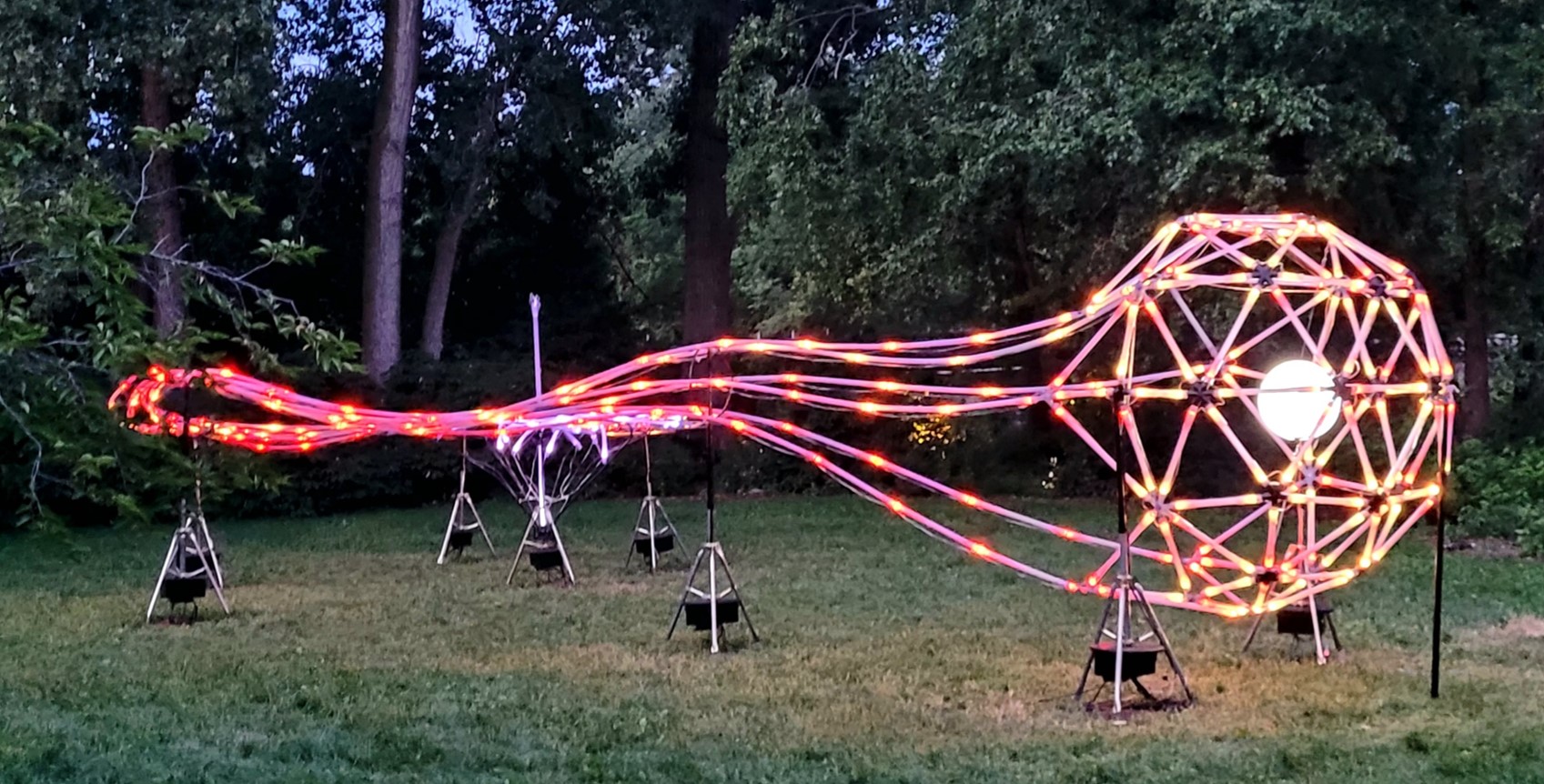}}\label{fig:fullcolor}}
    \caption{The sculpture at twilight showing the start of the light sequence, \textbf{(a)}, and at full color, \textbf{(b)}. }
    \end{figure}

\section{Installation}\label{install}
Installation of the work took six days. All the pieces were fabricated at the UW-Madison Physical Sciences Lab, and then loaded into a van and brought to Olbrich Gardens. Multiple people pitched in to ensure the project would be completed by opening night.  There was a great deal of "custom" wiring to get power and data to the LEDs in addition to physically joining the high quality waterproof connections on the LEDs. 

Once the mechanical and electrical components were installed, the software to control the sound and light was fine-tuned.  The final program was a sequence that lasted about 4 minutes that started by illuminating the star in white light that traveled down the "streams" to the swirl and be consumed by the black hole.  This was indicated by a change in color and sound. The beginning and middle of the color sequence is shown in 
Figure~\ref{fig:wlight} and Figure~\ref{fig:fullcolor}. 

\
 
\section{Response}\label{response}
\begin{figure}[h]
    \centering
     \subfloat[\centering]{{\includegraphics[width=6cm]{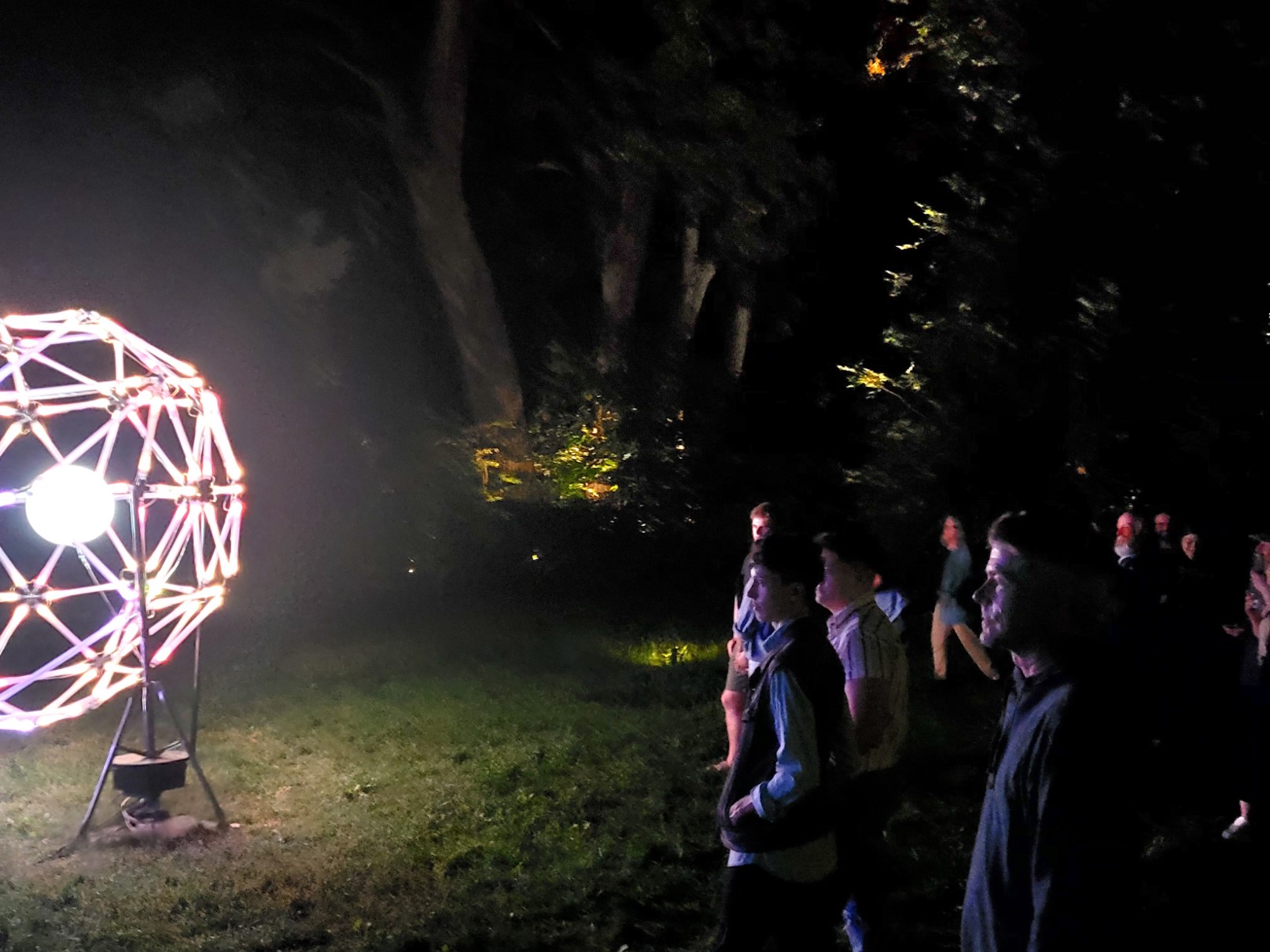}}\label{fig:starviewers}}
       \qquad
       \subfloat[\centering]{{\includegraphics[width=8cm]{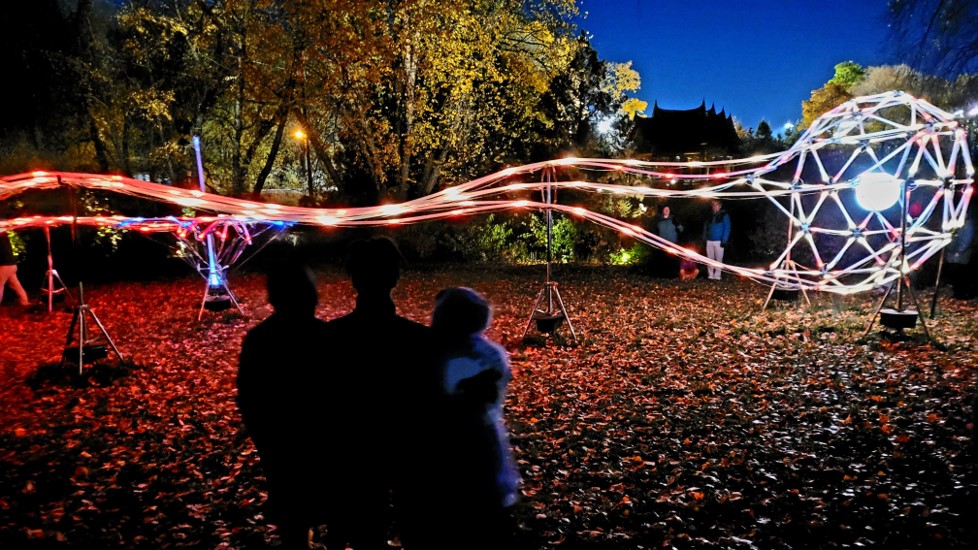}}\label{fig:fullviewers}}
    \caption{Viewers enraptured by the light sculpture, \textbf{(a)} and \textbf{(b)}. }
    \end{figure}

The TDE light sculpture was a tremendous hit, Figs,~\ref{fig:starviewers},~\ref{fig:fullviewers}.  People were intrigued by the sculpture with several interesting guesses of what it represented.  After reading the accompanying description or talking to volunteer docents assigned to the TDE sculpture, they were generally awed by the thought that a star could be pulled apart. As the pictures show, many people chose to view the sculpture from different vantage points, resulting in dramatic images and experiences,   
Figure~\ref{fig:starviewers}, Figure~\ref{fig:fullviewers}. 

The most common comment was an appreciation for the artistic interpretation depicting an almost incomprehensible event. The light display cycled through a 4 minute sequence that began with the sphere representing the star being illuminated with white lights. The star being pulled apart and consumed by the black hole was shown by illuminating the "arms" that streamed toward and eventually spiraled around the black sphere representing the black hole.  As the event progressed, the lights in the star sphere would fade until the last matter from the star was consumed.  This sequence was summarized on a plaque near the light sculpture.  A docent stationed at the light sculpture was available to answer questions as well. 

\section{Conclusion}\label{conclusion}
There are multiple ways to communicate science that go beyond expressing a technically challenging topic in an accessible language. We have developed strong partnerships with artists who bring different techniques, perspectives and often novel venues to connect with new audiences and communities.  This light sculpture is a prime example.  

\section{Key Contributions}\label{contributions}
We thank Faisal Abdu'Allah, for his help securing funding. We thank the WIPAC students and staff who helped with the fabrication, installation and tear down of the light sculpture---Vedant Basu, Delaney Butterfield, Ellen Bechtol, Hannah Haynes,  Jean DeMerit, Yuya Makino, and Laura Mercier.  Special thanks to Melissa Jeanne, Special Projects Manager at Olbrich Gardens, for her help during all stages of the project.

\bibliographystyle{ICRC}
\bibliography{references}

\providecommand{\href}[2]{#2}\begingroup\raggedright\begin{thebibliography}{10}

\bibitem{Madsen:20194t}
J.~Madsen, F.~Abdu'Allah, M.-D. Hosale, and M.~Ladoni
  \href{http://dx.doi.org/10.22323/1.358.0951}{{\em PoS} {\bfseries ICRC2019}
  (2019) 951}.

\bibitem{Madsen:2021D9}
J.~Madsen and L.~e. Mulot \href{http://dx.doi.org/10.22323/1.395.1381}{{\em
  PoS} {\bfseries ICRC2021} (2021) 1381}.

\bibitem{Rhino:McNeel}
R.~McNeel {\em et~al.} {\em Robert McNeel \& Associates, Seattle, WA} (2010) .

\bibitem{ShijiLighting}
``Shenzhen {S}hiji {L}ighting {C}o. {LTD}.''
\newblock \url{http://www.shiji-led.com/}.

\bibitem{Neopixels}
``{Neopixel Guide}.''
\newblock
  \url{https://learn.adafruit.com/adafruit-neopixel-uberguide/the-magic-of-neopixels/}.

\bibitem{Arduino}
A.~IDE, ``{Arduino IDE}.''
\newblock \url{https://www.arduino.cc/}.

\bibitem{Teensy}
``{Teensy 4.0}.''
\newblock \url{https://www.pjrc.com/store/teensy40.html/}.

\bibitem{OctoWS2811}
``{OctoWS2811 Adaptor}.''
\newblock \url{https://www.pjrc.com/store/octo28_adaptor.html/}.

\bibitem{OpenFrameworks}
``{OpenFrameworks}.''
\newblock \url{https://openframeworks.cc/}.

\bibitem{RaspberryPi}
``{Raspberry Pi 3 Model B+}.''
\newblock \url{https://www.raspberrypi.com/products/raspberry-pi-3-model-b/}.

\bibitem{SuperCollider}
``{SuperCollider}.''
\newblock \url{https://supercollider.github.io/}.

\end{thebibliography}\endgroup

%

\clearpage

\section*{Full Author List: IceCube Collaboration}

\scriptsize
\noindent
R. Abbasi$^{17}$,
M. Ackermann$^{63}$,
J. Adams$^{18}$,
S. K. Agarwalla$^{40,\: 64}$,
J. A. Aguilar$^{12}$,
M. Ahlers$^{22}$,
J.M. Alameddine$^{23}$,
N. M. Amin$^{44}$,
K. Andeen$^{42}$,
G. Anton$^{26}$
% ...
\\
\\
% ...
$^{12}$ Universit{\'e} Libre de Bruxelles, Science Faculty CP230, B-1050 Brussels, Belgium \\
% ...
$^{17}$ Department of Physics, Loyola University Chicago, Chicago, IL 60660, USA \\
$^{18}$ Dept. of Physics and Astronomy, University of Canterbury, Private Bag 4800, Christchurch, New Zealand \\
% ...
$^{22}$ Niels Bohr Institute, University of Copenhagen, DK-2100 Copenhagen, Denmark \\
$^{23}$ Dept. of Physics, TU Dortmund University, D-44221 Dortmund, Germany \\
% ...
$^{26}$ Erlangen Centre for Astroparticle Physics, Friedrich-Alexander-Universit{\"a}t Erlangen-N{\"u}rnberg, D-91058 Erlangen, Germany \\
% ...
$^{40}$ Dept. of Physics and Wisconsin IceCube Particle Astrophysics Center, University of Wisconsin{\textendash}Madison, Madison, WI 53706, USA \\
% ...
$^{42}$ Department of Physics, Marquette University, Milwaukee, WI, 53201, USA \\
% ...
$^{44}$ Bartol Research Institute and Dept. of Physics and Astronomy, University of Delaware, Newark, DE 19716, USA \\
% ...
$^{63}$ Deutsches Elektronen-Synchrotron DESY, Platanenallee 6, 15738 Zeuthen, Germany  \\
$^{64}$ Institute of Physics, Sachivalaya Marg, Sainik School Post, Bhubaneswar 751005, India \\
% ...

\subsection*{Acknowledgements}

\noindent
The authors gratefully acknowledge the support from the following agencies and institutions:
USA {\textendash} U.S. National Science Foundation-Office of Polar Programs,
U.S. National Science Foundation-Physics Division,
U.S. National Science Foundation-EPSCoR,
Wisconsin Alumni Research Foundation,
Center for High Throughput Computing (CHTC) at the University of Wisconsin{\textendash}Madison,
Open Science Grid (OSG),
Advanced Cyberinfrastructure Coordination Ecosystem: Services {\&} Support (ACCESS),
Frontera computing project at the Texas Advanced Computing Center,
U.S. Department of Energy-National Energy Research Scientific Computing Center,
Particle astrophysics research computing center at the University of Maryland,
Institute for Cyber-Enabled Research at Michigan State University,
and Astroparticle physics computational facility at Marquette University;
Belgium {\textendash} Funds for Scientific Research (FRS-FNRS and FWO),
FWO Odysseus and Big Science programmes,
and Belgian Federal Science Policy Office (Belspo);
Germany {\textendash} Bundesministerium f{\"u}r Bildung und Forschung (BMBF),
Deutsche Forschungsgemeinschaft (DFG),
Helmholtz Alliance for Astroparticle Physics (HAP),
Initiative and Networking Fund of the Helmholtz Association,
Deutsches Elektronen Synchrotron (DESY),
and High Performance Computing cluster of the RWTH Aachen;
Sweden {\textendash} Swedish Research Council,
Swedish Polar Research Secretariat,
Swedish National Infrastructure for Computing (SNIC),
and Knut and Alice Wallenberg Foundation;
European Union {\textendash} EGI Advanced Computing for research;
Australia {\textendash} Australian Research Council;
Canada {\textendash} Natural Sciences and Engineering Research Council of Canada,
Calcul Qu{\'e}bec, Compute Ontario, Canada Foundation for Innovation, WestGrid, and Compute Canada;
Denmark {\textendash} Villum Fonden, Carlsberg Foundation, and European Commission;
New Zealand {\textendash} Marsden Fund;
Japan {\textendash} Japan Society for Promotion of Science (JSPS)
and Institute for Global Prominent Research (IGPR) of Chiba University;
Korea {\textendash} National Research Foundation of Korea (NRF);
Switzerland {\textendash} Swiss National Science Foundation (SNSF);
United Kingdom {\textendash} Department of Physics, University of Oxford.

\end{document}